\documentclass[aps,prb,preprint,unsortedaddress,superscriptaddress]{revtex4-1} 

\usepackage{amsmath}  
\usepackage{amsfonts} 
\usepackage{graphicx} 
\usepackage{subfigure}
\usepackage[T1]{fontenc}
\usepackage{color}

\begin{document}


\title{Kinetics of spontaneous filament nucleation via oligomers: insights from theory and simulation}

\author{An\dj ela~\v{S}ari\'c\normalfont\textsuperscript{a}}
\email{a.saric@ucl.ac.uk}
\thanks{\normalfont\textsuperscript{a}equal contribution}
\affiliation{Department of Physics and Astronomy, Institute for the Physics of Living Systems, University College London, Gower Street, London, WC1E 6BT, U. K.}
\affiliation{Department of Chemistry, University of Cambridge, Lensfield Road, Cambridge, CB2 1EW, U.K.}

\author{Thomas C. T. Michaels\normalfont\textsuperscript{a}}
\email[]{tctm3@cam.ac.uk}
\thanks{\normalfont\textsuperscript{a}equal contribution}
\affiliation{Department of Chemistry, University of Cambridge, Lensfield Road, Cambridge, CB2 1EW, U.K.}
\affiliation{Paulson School of Engineering and Applied Sciences, Harvard University, Cambridge, MA 02138, USA.}

\author{Alessio Zaccone}
\affiliation{Department of Chemical Engineering, University of Cambridge, Pembroke St, Cambridge CB2 3RA, U.K.}

\author{Tuomas P. J. Knowles}
\affiliation{Department of Chemistry, University of Cambridge, Lensfield Road, Cambridge, CB2 1EW, U.K.}

\author{Daan Frenkel}
\affiliation{Department of Chemistry, University of Cambridge, Lensfield Road, Cambridge, CB2 1EW, U.K.}

\begin{abstract}
Nucleation processes are at the heart of a large number of phenomena, from cloud formation to protein crystallization. A recently emerging area where nucleation is highly relevant is the initiation of filamentous protein self-assembly, a process that has broad implications in many research areas ranging from medicine to nanotechnology. As such, spontaneous nucleation of protein fibrils has received much attention in recent years with many theoretical and experimental studies focussing on the underlying physical principles. In this paper we make a step forward in this direction and explore the early time behaviour of filamentous protein growth in the context of nucleation theory. We first provide an overview of the thermodynamics and kinetics of spontaneous nucleation of protein filaments in the presence of one relevant degree of freedom, namely the cluster size. In this case, we review how key kinetic observables, such as the reaction order of spontaneous nucleation, are directly related to the physical size of the critical nucleus. We then focus on the increasingly prominent case of filament nucleation that includes a conformational conversion of the nucleating building-block as an additional slow step in the nucleation process. Using computer simulations, we study the concentration dependence of the nucleation rate. We find that, under these circumstances, the reaction order of spontaneous nucleation with respect to the free monomer does no longer relate to the overall physical size of the nucleating aggregate but rather to the portion of the aggregate that actively participates in the conformational conversion. Our results thus provide a novel interpretation of the common kinetic descriptors of protein filament formation, including the reaction order of the nucleation step or the scaling exponent of lag times, and put into perspective current theoretical descriptions of protein aggregation.
\end{abstract}

\maketitle

\section{Introduction}
Nucleation is the initial step in the formation of a new ordered structure through self-organization. It is characterized by the presence of a free energy barrier to form the smallest growth-competent unit of the new structure. Many phenomena in nature, science and engineering are nucleated processes, including everyday examples such as cloud formation, ice crystallization, the boiling of water, or the formation of bubbles in a champagne glass. A particularly intriguing example of a nucleated process is the formation of protein filaments, which is the topic of the present paper. This is a fundamental form of biological self-assembly with important implications in areas ranging from medicine to materials science. Biofilaments of actin and tubulin for instance underlie key events in cellular life, such as providing the rigidity of the cellular cytoskeleton or participating in cell motility and cell division.\cite{Alberts,Oosawa_1975,Oosawa_1962,Tobacman_1983, Frieden_1983, Cooper_1983, Frieden_1985}  
On the other side, aberrant filamentous protein aggregation is associated with over 50 increasingly prevalent human disorders, such as Alzheimer's, Parkinson's diseases and type II diabetes.\cite{Dobson_2002,Dobson_2003,Chiti_2006,Selkoe_2007,Rhoades_2000,Dauer_2003,DiFiglia_1997} These pathologies are intimately associated with the formation and deposition in the brain or other organs of fibrillar protein aggregates, commonly known as  amyloids, which are the result of the aggregation of normally soluble and functional proteins into elongated fibrillar structures characterized by their $\beta$-sheet rich structure. Amyloids, however, are not only associated with disease, as it was initially believed, but have been increasingly found to serve also many functional roles within living organisms\cite{Fowler_2007} and this natural use of the amyloid state of proteins and peptides for functional purposes has inspired many applications of these structures as materials for nanotechnology.\cite{Reches_2003,Knowles_2010,Knowles_2011b,Gazit_2007,Li_2012} The formation of protein filaments has been established to be a nucleated polymerization process where a slow spontaneous fibril nucleation step, also referred to as primary nucleation, is followed by rapid growth through filament elongation\cite{Oosawa_1975,Hofrichter_1974,Ferrone_1985b,Ferrone_1980,Bishop_1984,Hofrichter_1986,Jarrett_1993,Lomakin_1997,Harper_1997,Galkin_2004,Serio_2000,Nguyen_2004,Nguyen_2005,Nguyen_2006,Ferrone_2006,Galkin_2007,Linse_2007,Auer_2007,Auer_2008,Xue_2008,Zhang_2009,Knowles_2009} and, in certain cases, self-replication through secondary pathways.\cite{Cabaleiro-Lago_2008,Linse_2007,Ferrone_1980,Cohen_2013,Ruschak_2007,frenkel2016}
Here, the term ``spontaneous'' refers to the fact that the random formation of the smallest growth-competent aggregates (nuclei) occurs directly from solution, without the participation of surfaces or nucleation seeds.
A particularly useful approach to understand the way in which soluble proteins are converted into their fibrillar counterparts through spontaneous nucleation is represented by kinetic models of filamentous growth\cite{Cohen_2011a,Cohen_2011b,Cohen_2011c,Cohen_2013,Knowles_2009,Oosawa_1975}. These kinetic models allow the underlying molecular-level mechanisms of fibril formation to be connected with {\it{in vitro}} experimental measurements of the aggregate mass concentration e.g.~by fluorescence microscopy or related techniques.\cite{Hellstrand_2010} In these models, the spontaneous fibril nucleation step is commonly described as an $n_c$-th order reaction with respect to the free monomer concentration $c$ with rate
\begin{equation}
r = k_n c^{n_c},
\end{equation}
where $k_n$ is the rate constant for spontaneous nucleation and $n_c$ is an effective reaction order of spontaneous nucleation. Because nucleation is slow compared to growth, the value of $n_c$ can be obtained experimentally from the concentration dependence of the half-polymerization time $t_{1/2}$ (defined as the time at which half of the monomers mass is sequestered in aggregates), as the slope of this relationship in a double logarithmic plot gives the so called scaling exponent $\gamma$ defined by:
\begin{equation}
t_{1/2} = A \ c^{\gamma},\quad \gamma =-n_c/2.
\end{equation}
A key question, in the field, is how to relate these experimentally measured kinetic descriptors, including the reaction order $n_c$ and the scaling exponent $\gamma$, with the microscopic characteristics of the underlying nucleation step, such as the physical size of the nucleating aggregates. This connection provides important insights into the nature of the nucleation process from experimental measurements. This problem has already received significant attention in the protein aggregation literature, and we start here by providing a brief overview of the simplest case of fibril nucleation by direct polymerization of monomers, incorporating the published theories and quantitative experiments.

We then make a step forward and consider the increasingly evident process of protein nucleation which includes a conformational change of the nucleating protein, giving rise to multi-step nucleation processes via small oligomers. We study the kinetics of such a process using coarse-grained computer simulations, and provide a novel physical interpretation of the related kinetic parameters that are commonly measured in experiments. In particular, we investigate the physical interpretation of the reaction order $n_c$ when proteins undergo a conformational change during nucleation, and find that $n_c$ is determined by the portion of the oligomer size that directly participates in the conformational conversion step.

\section{Kinetics of spontaneous fibril nucleation with one degree of freedom}
We start our discussion by reviewing the simplest model of fibril nucleation, in which aggregates of different sizes but same structure are formed by direct polymerization of protein monomers (Fig. 1(a)). We demonstrate that this model gives rise to spontaneous nucleation if the cluster free energy function has a maximum as a function of cluster size. Under these circumstances, it is found that, independently of the specific form of the cluster free energy function, the reaction order of spontaneous nucleation $n_c$ is linked to the number of monomers that compose the fibril nucleus.\cite{Kashchiev_1982} The simplicity of this model arises from the fact that the aggregate size is the only degree of freedom in the system; in Section \ref{sec_conv} we relax this assumption by considering the effect of other potentially important degrees of freedom, such as the internal structure of clusters.

\subsection{Direct polymerization models of spontaneous fibril nucleation and the nucleation theorem}

To see how spontaneous nucleation emerges from a direct polymerization model, we consider the following master equation describing the time evolution of the concentration $f(t,N)$ of aggregates of $N$ monomers under the action of elongation and dissociation processes:\cite{footnote_BD, Becker_1935, Penrose, Zeldovich_1942, Zaccone_2013}
\begin{align}\label{eq_master}
\frac{\partial f(t,N)}{\partial t} & = c\ k_{\textrm{on}}(N-1)  f(t,N-1)-c\ k_{\textrm{on}}(N)  f(t,N)\\\nonumber
& \quad +k_{\textrm{off}}(N+1) f(t,N+1)-k_{\textrm{off}}(N) f(t,N),
\end{align}
where $c$ is the free monomer concentration, $k_{\textrm{on}}(N)$ and $k_{\textrm{off}}(N)$ are the (size-dependent) rate constants for the addition and removal of monomers. These rate constants are linked together by the detailed balance condition 
\begin{equation}\label{db}
c\ k_{\textrm{on}}(N-1) f_{\textrm{eq}}(N-1)= k_{\textrm{off}}(N) f_{\textrm{eq}}(N)
\end{equation}
where $f_{\textrm{eq}}(N)$ is the cluster size distribution function at equilibrium.
As detailed in Refs.~\cite{Zaccone_2013,Zeldovich_1942}, Eqs.~\eqref{eq_master} can be mapped onto a one-dimensional diffusion equation in a potential landscape by assuming that $f(t,N)$ varies sufficiently smoothly with $N$ so that the continuum limit approximation applies: 
\begin{equation}\label{diff_eq}
\frac{\partial f(t,N)}{\partial t} = k_{\textrm{on}}c\ \frac{\partial }{\partial N}\left[\frac{\partial f(t,N)}{\partial N} +\beta \frac{\partial \Phi(N)}{\partial N} f(t,N)\right],
\end{equation}
where $\beta=1/(kT)$ denotes the inverse temperature ($k$ is the Boltzmann constant) and we have introduced the cluster free energy function $\Phi(N)$ defined by the relationship
\begin{equation}
\frac{f_{\textrm{eq}}(N)}{c}= e^{-\beta \Phi(N)}.
\end{equation}
If the free energy function $\Phi(N)$ has a maximum at some value $N^*$, corresponding to the critical nucleus size, then an expression for the rate of nucleation can be obtained from Eq.~\eqref{diff_eq} using the saddle point approximation:\cite{Zeldovich_1942, Zaccone_2013,Kramers_1940}
\begin{equation}\label{rate_kram}
r \sim k_{\textrm{on}}c\ \left(\frac{\Phi''(N^*)}{2\pi\beta^{-1}}\right) e^{-\beta \Phi(N^*)}.
\end{equation}
Thus, the master equation \eqref{eq_master} of direct polymerization yields spontaneous nucleation. 
An important point to recognize here is that Eq.~\eqref{rate_kram} is valid for arbitrary cluster free energy functions, so that depending on the specific form of $\Phi(N)$ several models of spontaneous nucleation can be formulated.
Classical nucleation theory,\cite{Abraham_1974,Kashchiev_2000} for instance, describes clusters of dimensionality $d$ as an object with associated volume and surface energy terms:
\begin{equation}\label{3d}
\Phi(N) = a \sigma N^{\frac{d-1}{d}}- N\Delta\mu, \quad (d>1) 
\end{equation}
where $\Delta \mu=\beta^{-1}\log(c/c_s)$ is the supersaturation, $c_s$ the saturation concentration, $\sigma$ the surface tension (energy per unit surface) of the interface between the aggregate and the surrounding solvent and $a$ is a geometrical prefactor. 
The balance between unfavourable entropy contribution from the loss of molecular degrees of freedom and the favourable energy from the bonds between monomers creates a barrier. As $N$ increases, more bonds are created between monomers eventually overcoming the unfavourable entropy contributions that make small cluster unstable. 
The nucleus formation is the rate limiting step and corresponds to the point at which the free energy $\Phi(N)$ peaks (Fig.1(b)), and according to Eq.~\eqref{rate_kram} the rate of nucleation is given by:
\begin{equation}\label{rate_eq_CNT}
r \sim c \ e^{-\frac{\beta \Delta\mu N^*}{d-1}},\quad N^* =\left(\frac{(d-1)a\sigma}{d\ \Delta \mu }\right)^d .
\end{equation}
It is easy to verify from Eq.~\eqref{rate_eq_CNT} that the nucleus size $N^*$ satisfies the relationship:
\begin{equation}\label{theorem}
N^* = \frac{d \log(r)}{d \log(c)}-1= n_c-1,
\end{equation}
where the factor $-1$ comes from the concentration dependence of the prefactor in Eq.~\eqref{rate_kram}. Equation \eqref{theorem} is a key result: it states that the nucleus size can be obtained from the slope of a double logarithmic plot of the nucleation rate $r$ against the monomer concentration $c$ and provides a direct relationship linking the physical size of nuclei to the experimentally accessible reaction order $n_c$ of spontaneous nucleation. Importantly, Eq.~\eqref{theorem}, which was derived here for the specific cluster free energy function of Eq.~\eqref{3d}, turns out to be a far more general result known as the nucleation theorem.\cite{Kashchiev_1982} This theorem states that Eq.~\eqref{theorem} is valid for arbitrary cluster free energy functions of the form $\Phi(N) = F(N)- N\Delta\mu $, so that in a nucleating system where the aggregate number is the only relevant degree of freedom, the experimental kinetic parameters $n_c$ and $\gamma$ can always be linked directly to the physical size of the nuclei. As a final note, we remind here that, in the context of classical nucleation theory, the surface energy term $F(N)$ of one dimensional clusters ($d=1$) is independent of $N$ so that the cluster free energy function $\Phi(N)$ has no maximum (Fig.~1(b)). Hence, direct polymerization in 1D is a downhill process, where every aggregate is more stable that the previous ones; there is no classical nucleation in a truly 1D system.\cite{Kashchiev_2000}

\begin{figure}
\includegraphics[width=\columnwidth]{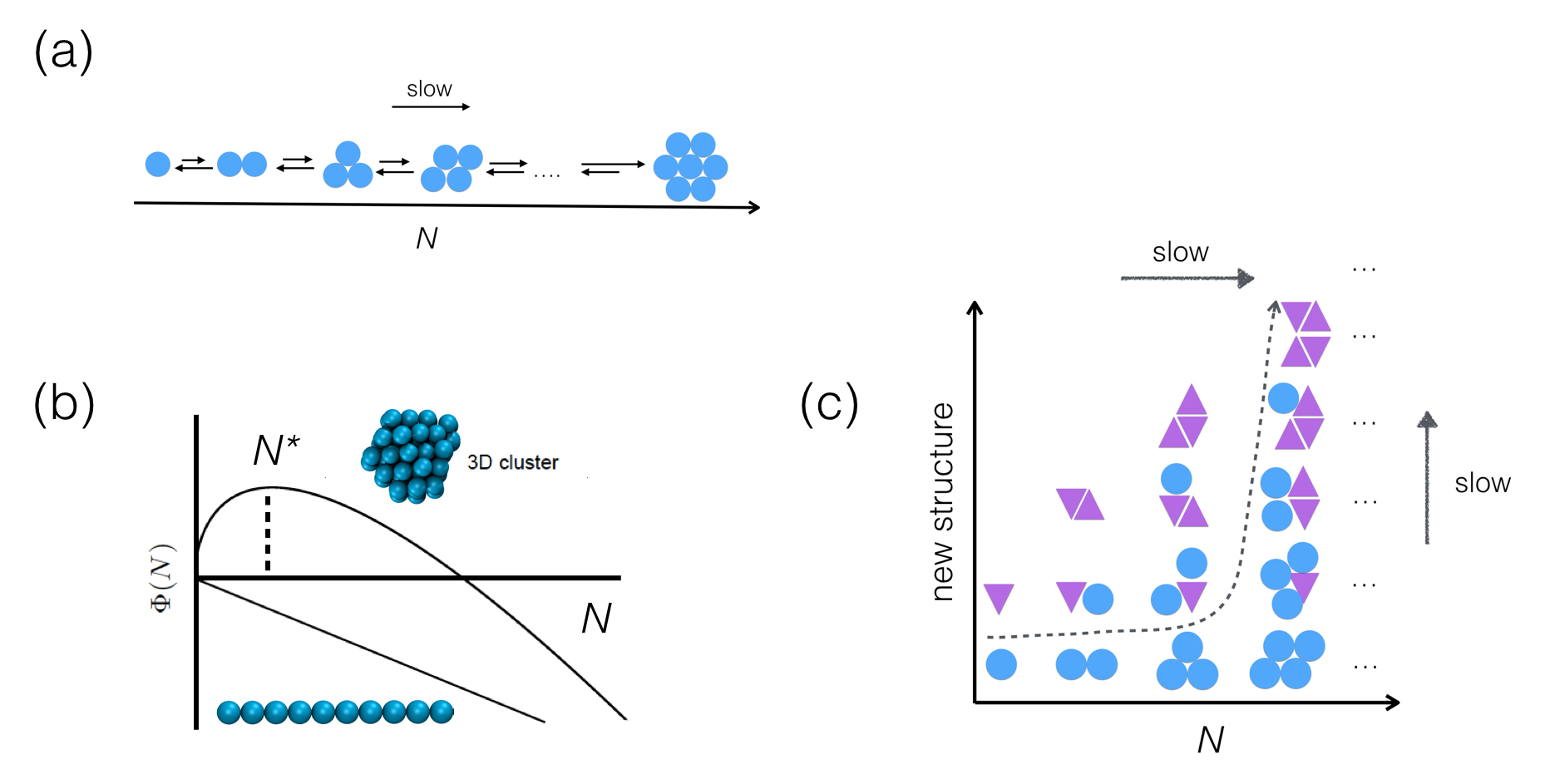}
\caption{\label{figure_CNT} (a)  One-step nucleation is characterised by a single slow coordinate that corresponds to the cluster size $N$. (b)  Cluster free energy of classical nucleation theory for a 3D spherical cluster and for a 1D cluster. Note that in 1D there is no free energy barrier. (c) Multistep nucleation is characterised by additional slow coordinates. In the example shown here, the additional slow coordinate corresponds to a structural change. The dashed line indicates a possible nucleation pathway.}
\end{figure}

Direct polymerization models have been used widely in the protein aggregation literature to describe spontaneous fibril nucleation. Important examples include classical nucleation theory descriptions of amyloid as elongated 2D crystals.\cite{Kashchiev_2010,Cabriolu_2010,Cabriolu_2011} By considering prismatic aggregates of fixed width but variable length and thickness build up by successively layered $\beta$-sheets, expressions for the nucleation rate, nucleus size and nucleation work have been obtained. In accordance with the nucleation theorem Eq.~\eqref{theorem}, these theories predict the existence of a well-defined nucleus size which can be obtained from a log-log plot of the nucleation rate against monomer concentration.
Recent simulation and theoretical studies of fibril nucleation have highlighted that introducing an interaction anisotropy in these models results in a non-standard nucleation mechanism where the concept of a well-defined nucleus size breaks and nuclei with varying size are observed instead.\cite{Cabriolu_2012,Kashchiev_2013,Bingham_2013}

\subsection{Pre-equilibrium kinetic models of spontaneous fibril nucleation}

Worth mentioning at this point are pre-equilibrium kinetic models of fibril nucleation, which often correspond to direct polymerization mechanisms. These models are characterized by two main assumptions. The first assumption is that clusters do not possess internal structure, so that the nucleus can be considered as a small piece of a long aggregate and the cluster size is the only relevant degree of freedom in the system. The second assumption is that the nuclei are in equilibrium with the soluble monomer, so that their concentration can be obtained by equating the respective chemical potentials and the original kinetic problem is now transformed into an equilibrium one. In these models the rate of nucleation can be summarized as:\cite{Ferrone_1980,Bishop_1984,Ferrone_1985b,Ferrone_1999,Vitalis_2011,Oosawa_1975,Oosawa_1962,Knowles_2009,Cohen_2011a, Andrews_2007,Hofrichter_1974}
\begin{equation}
r= A \ c^{n_c}.
\end{equation} 
The exact connection between the exponent $n_c$ and the nucleus size is model dependent. For example, some authors\cite{Ferrone_1980,Bishop_1984,Ferrone_1985b,Ferrone_1999,Vitalis_2011} describe the rate of nucleation as the rate at which nuclei elongate $r=k_+c\ [N^*]$, where $k_+$ is the elongation rate constant and $[N^*]$ is the concentration of nuclei. The assumption that nuclei are in equilibrium with the monomers results in $[N^*]= K_{n}c^{N^*}$, where $K_n$ is the equilibrium constant for the nucleus-monomer equilibrium. Combining these expressions yields $r=k_+K_n c^{N^*+1}$ and we find the following relationship between the reaction order $n_c$ of spontaneous nucleation and $N^*$\cite{Kashchiev_1982}
\begin{equation}\label{Ncrit_2}
N^* = \frac{d\log (r)}{d\log(c)}-1=n_c-1.
\end{equation}
Other authors have employed similar pre-equilibrium arguments to obtain alternative expressions for the nucleation rate that differ in the interpretation of the reaction order, including $N^*=n_c+1$ or $N^*=n_c$, but not in substance \cite{Hofrichter_1974, Oosawa_1975,Oosawa_1962,Knowles_2009,Cohen_2011a, Andrews_2007}. In all these models, in fact, the value of the nucleus size is directly related to the slope of a plot of $\log(r)$ against the logarithm of the monomer concentration, and thus the physical size of nuclei can in principle be accessed from kinetic measurements. Several variations of these pre-equilibrium models of nucleation have also been formulated, for instance by considering different addition rates for monomer above and below $N^*$,\cite{Ferrone_1999,Bishop_1984,G102,G113,G118} by assuming that the nucleus is formed through successive associations of small oligomers\cite{G104,G116,G124,Flyvbjerg_1996b} or by including reversible association steps for aggregate sizes below $N^*$ and considering only irreversible polymerization for $N>N^*$.\cite{G71,G104,G116,G124}

\section{Kinetics of spontaneous fibril nucleation with multiple degrees of freedom}\label{sec_conv}

So far, we have considered the situation when the relevant degree of freedom of the nucleating system is the physical size of aggregates. Under these circumstances, the nucleation theorem\cite{Kashchiev_1982} provides a direct relationship between the nucleus size and the concentration dependence of the nucleation rate. This result offers a powerful strategy for accessing key information about the underlying nucleation step from experimental measurements of aggregation kinetics. We now make a step forward and consider the more complicated situation when the nucleation process is controlled by additional relevant degrees of freedom, such as the internal structure of clusters. A prominent example of such a situation is realized when the aggregating species change their shape or conformation during nucleation. In fact, when considering the assembly of soft species, such as proteins, one needs to take into account the fact that the species within the final aggregate might be in a significantly different state from their counterparts in solution~\cite{chung2010self,shin2012direct}. This situation can be viewed as a nucleation process that is governed by multiple degrees of freedom (e.g. molecular rearrangements), in addition to the cluster size, as sketched in Fig.~1(c). An important realization of this scenario is the aggregation of amyloid fibrils, where proteins acquire a $\beta$-sheet conformation within the fibril, which is typically very different from their native conformations in solution. A large number of structurally unrelated proteins form this type of fibrils, hence amyloid fibril formation is regularly accompanied by a marked change in protein conformation. As demonstrated below, in this scenario we do not find a single-valued relationship between nucleus size $N^*$ and the reaction order $n_c$, but rather find that for a given value of $n_c$ the overall size of the nucleating oligomers can change. Interestingly, however, we find that a modified nucleation theorem connects $n_c$ to the sub-oligomer size within which the conformational change takes place.

\subsection{Conformational change in amyloid nucleation}
Amyloidogenic proteins can be characterised according to their propensity to change their conformation and acquire the $\beta$-sheet prone structure~\cite{Mayo_1999, Tjernberg_2002, Pawar_2005, Zibaee_2007}. This propensity for the $\beta$-sheet controls the rate and pathways of amyloid fibril nucleation ~\cite{caflisch06, caflisch07, shea09, shea09_1, caflisch10, shea11, shea15}. For proteins with low $\beta$-sheet propensity,  the conformational change from the native into the $\beta$-sheet form is slow and energetically unfavourable, and the fraction of proteins in the $\beta$-sheet conformation can act as a second slow degree of freedom in amyloid fibril formation, in addition to the aggregation number.\par
A number of experimental studies have investigated amyloid nucleation pathways and reported the existence of non-$\beta$-sheet clusters during amyloid formation~\cite{Garzon_1997,
Shankar_2008,Liang_2010,Bernstein_2005,Bitan_2003,Bleiholder_2011,Sabate_2005,Yong_2002,cho2011multistage}, suggesting a multi-step nucleation mechanisms, where fibril nucleation takes place via disordered prefibrillar clusters.\cite{Lomakin_1997,Serio_2000,Galkin_2007,Hellstrand_2010,Vekilov_2010,Lee_2011} This nucleation scenario, also called a nucleated conformational conversion or a two-step nucleation, has also been in focus of several theoretical studies~\cite{Kashchiev_2005,caflisch06, caflisch07, shea09_1,Vitalis_2010,Schmit_2011,Auer_2012,frenkel2014, groot2016,switch2016,Briels2016}.\par

\subsection{Computer simulations of amyloid nucleation}
As a quantitative understanding of aggregation processes with more than one slow degree of freedom is still presently lacking, it is beneficial to obtain interpretation of its kinetic descriptors. Coarse-grained computer simulations can be of great help in this case. Here, we use coarse-grained Monte Carlo simulations to study the kinetics and thermodynamics of nucleation of amyloid-like fibrils, for proteins with a range of $\beta$-propensities, attempting to rationalise experimentally measurable kinetic parameters in terms of the underlying microscopic steps. Amyloid fibril formation is known to involve pre-nucleation disordered oligomers, which do not possess much $\beta$-sheet content typical for fibrils, hence the oligomer aggregation number emerges as one slow degree of freedom, while the $\beta$-sheet content, typical for fibrils, emerges as the second one. Our model, although presented in the context of amyloid nucleation, is generic and can be applied to formation of any protein filaments which involves a conformational change.\par

We employ a minimal Monte Carlo computational model that reproduces fibril nucleation, as described in our previous work~\cite{frenkel2014,frenkel2016}. Briefly, the model accounts for the fact that amyloidogenic peptides and proteins exist in minimally two states: a state in solution (denoted ``$s$'') that can form disordered oligomers, and a higher free-energy state that can form the $\beta$-sheet enriched fibrils (denoted ``$\beta$'')~\cite{vacha,frenkel2014}. The ``$s$'' state is modelled as a hard spherocylinder with an attractive patch at the tip, which represents non-specific interprotein interactions, and drives the formation of small disordered oligomers, as depicted in Fig. 2 (a) and (b). The strength of the attraction between two ``$s$'' proteins is characterised with a parameter $\epsilon_{ss}$, set to $\epsilon_{ss}=5.5\textrm{k}T$ in this paper. The fibril forming configuration is described as a hard spherocylinder with an attractive side patch, which captures the interactions between the $\beta$-sheet prone state, such as the hydrogen-bonding and hydrophobic interactions, and leads to the fibrillar aggregates (Fig. 2 (a) and (b)). The magnitude of this attraction is considerably stronger than that between proteins in the soluble state, with $\epsilon_{\beta\beta}=30\textrm{k}T$. The ``$s$'' -``$\beta$'' interaction was set to $\epsilon_{s \beta}=\epsilon_{ss} +1\textrm{k}T$, as in our previous work~\cite{frenkel2014}. Throughout the text $\textrm{k}$ denotes Boltzmann's constant and $T$ the temperature.\par
We start our simulations with $600$ proteins randomly distributed in a periodic cubic box, with all proteins in the ``$s$'' state. A protein is randomly chosen to be swapped between the ``$s$''  and ``$\beta$'' state with a probability $P_{swap}$. The ``$s$''  $\rightarrow$ ``$\beta$''  swap is thermodynamically unfavourable, and is penalised with an excess chemical potential of $\Delta\mu_{s\beta}$, to reflect the loss of the conformational entropy of the $\beta$-hairpin compared to the form in solution. This value of $\Delta\mu_{s\beta}$ quantifies the protein's  $\beta$-sheet propensity, and controls the additional slow degree of freedom. The degree of oligomerisation of the proteins in the ``$s$''-state can be controlled via the protein concentration $c$, as probed in the text. ~\cite{footnote_OL}\\

To obtain information about the kinetics of fibril nucleation, we use the mean first passage time as the proxy for nucleation rate, and calculate the rate of primary nucleation as the inverse of the average lag time for nucleation~\cite{frenkel2016}. The lag time is defined as the number of MC steps needed for the first oligomer containing at least two $\beta$-proteins to appear in the simulation, since the appearance of such a nucleus always leads to further fibril growth in our simulations. We however note that such an oligomer can contain any composition of the proteins in the  ``$s$''  state and can be of any overall size. The average lag time is then calculated from at least $6$ repetitions of the same system with different random seeds, and is expressed in the units of $10^8$ MC steps.\par

\begin{figure*}[t!]
\centering
\includegraphics[width=0.75\textwidth]{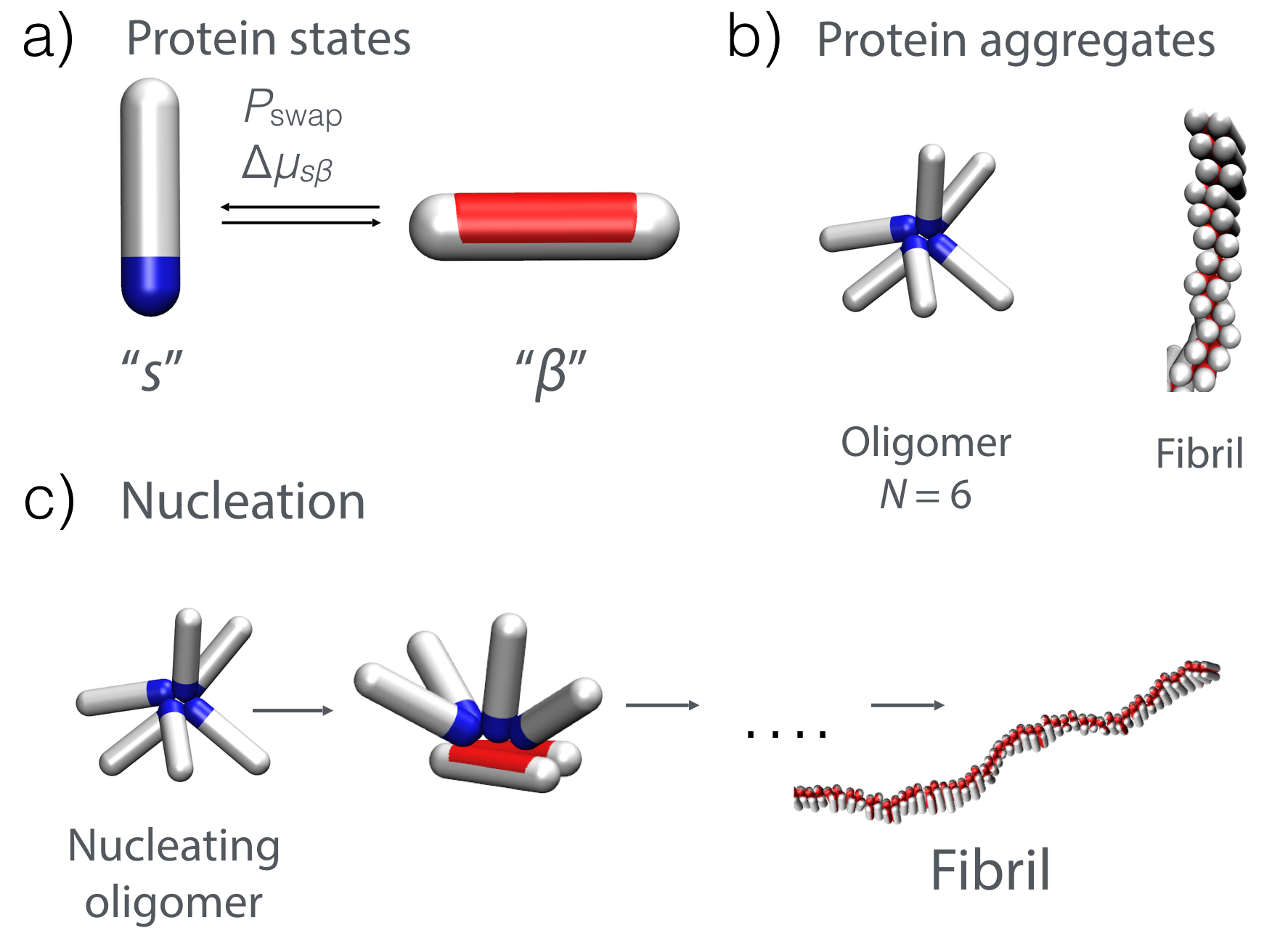}\label{fig:Fig2}
\caption{The Monte Carlo model. (a) The protein can switch between two states: the soluble state, ``$s$'', that is lower in energy, interacts weakly with its own kind, and forms disordered oligomers; and the $\beta$-sheet prone state,``$\beta$'', which is higher in energy, but interacts stronger with its own type than the soluble state, and forms fibrils. Attractive patches are coloured in blue and red for the  ``$s$'' and  ``$\beta$'' state respectively. (b) Aggregates of the two possible states: disordered oligomer (left panel) and fibril (right panel). (c) At low protein concentrations nucleation proceeds via oligomers. Illustration of an oligomer of the size $N=6$, where the nucleation takes place when two proteins simultaneously convert into the $\beta$-prone state, which triggers the nucleus growth into a long fibril.}
\end{figure*}

\subsection{Kinetics of amyloid nucleation computed in simulations}
At low protein concentrations, which is the regime we focus on in this paper, nucleation proceeds via oligomers, as depicted in Fig. 2(c).  Starting from an equilibrated population of soluble proteins and their oligomers, we have calculated the rate of fibril nucleation across a wide range of protein concentrations, as shown in Fig. 3(a), for the case of a protein with a low propensity for the $\beta$-state  ($\Delta \mu_{s\beta}=20\textrm{k}T$) and the protein with a relatively high $\beta$-propensity ($\Delta \mu_{s\beta}=10\textrm{k}T$). As expected, the rate of nucleation increases with increasing protein concentration, and protein $\beta$-propensity. The scaling exponent, $n_c$, which relates the reaction order of the nucleation step with the monomer concentration is given by the slope of the plot in Fig. 3(a), and is found to decrease with increasing $\beta$-propensity.\\
What is the physical interpretation of the scaling exponent $n_c$? An obvious characteristic of $n_c$ is that it decreases sharply in the vicinity of the critical micelle concentration (\textit{cmc}). The \textit{cmc} is the concentration above which increasing the total protein concentration leaves the concentration of monomers in solution unaffected, causing the weaker dependence of nucleation the rate on the protein concentration observed above the \textit{cmc} (Fig. 3(b)). 
Since amyloidogenic proteins typically occur at low concentrations, as low as nanomolar, we focus here on the meaning of the exponent at low concentrations, much before the \textit{cmc} is reached. In this concentration regime, the measured scaling exponent is $n_c \approx 4.5$ and $\gamma \approx 2$, for the protein with a low and high $\beta$-propensity respectively (Fig. 3(a)) .\\
In light of the various theories of nucleation discussed in the Section II, where the scaling of the nucleation rate was found to be linked with the protein concentration raised to the critical nucleus size (Eq.(10)), we studied the size (aggregation number) of the nucleating oligomers, a parameter that can be accessed directly in our simulations. Over the concentration range studied here, the average aggregation number of the nucleating oligomer changes between $N^* \approx 2- 12$ for the low $\beta$-propensity protein (red circles in Fig. 3(c)) and between $N^* \approx 2 - 4$ for the high $\beta$-propensity one (blue crosses in Fig. 3(d)). Clearly, the size of the nucleating oligomer increases with the increase in the protein concentration, as predicted in our previous analysis of free energy barriers for nucleation via small oligomers~\cite{frenkel2014}. However, the size of the nucleating oligomer does not correspond to the value of the reaction order, and this discrepancy becomes more prominent for the protein with low $\beta$-propensity. Instead, we find that the reaction order corresponds to the subset of proteins in the oligomer that directly participate in the conversion step, as shown by the black filled circles in Fig. 3(c) for the low $\beta$-propensity protein, and the black squares in Fig. 3(d) for the high $\beta$-propensity protein. This subset includes the converting protein, and the proteins that directly interacted with it in the conversion step. Not unexpectedly, the conformational conversion ceases to be a slow degree of freedom for the high $\beta$-propensity protein, and the situation where the reaction order corresponds to the oligomer size, described by classical nucleation, is recovered.\\
At this point, it is worth noticing that in order to test for the presence of possible additional time-scales involved in our results, we repeated the rate measurements for a larger value of the  ``$s$''  $\rightarrow$ ``$\beta$'' conversion attempt, $P_{swap}=1$ in our MC scheme (data not shown). In this case, the reaction was overall faster, but the scaling exponent and sizes of the nucleating oligomers remained unchanged, as the free energy landscape of the system remained unaltered. 

 \begin{figure*}[t!]
\centering
\subfigure[]{\includegraphics[width=0.45\textwidth]{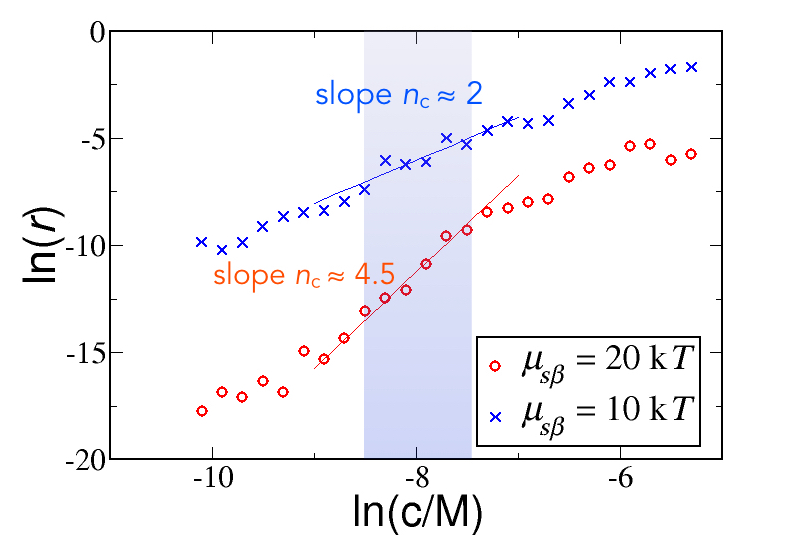}}
\subfigure[]{\includegraphics[width=0.45\textwidth]{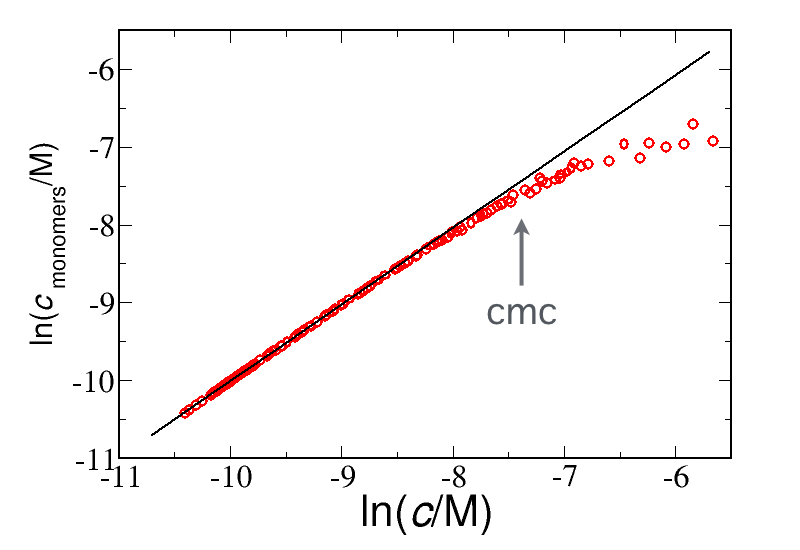}}
\subfigure[]{\includegraphics[width=0.45\textwidth]{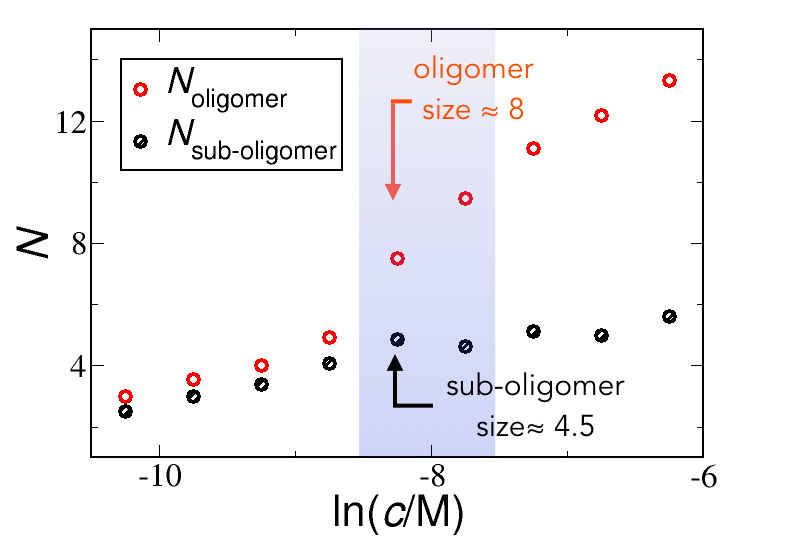}}
\subfigure[]{\includegraphics[width=0.45\textwidth]{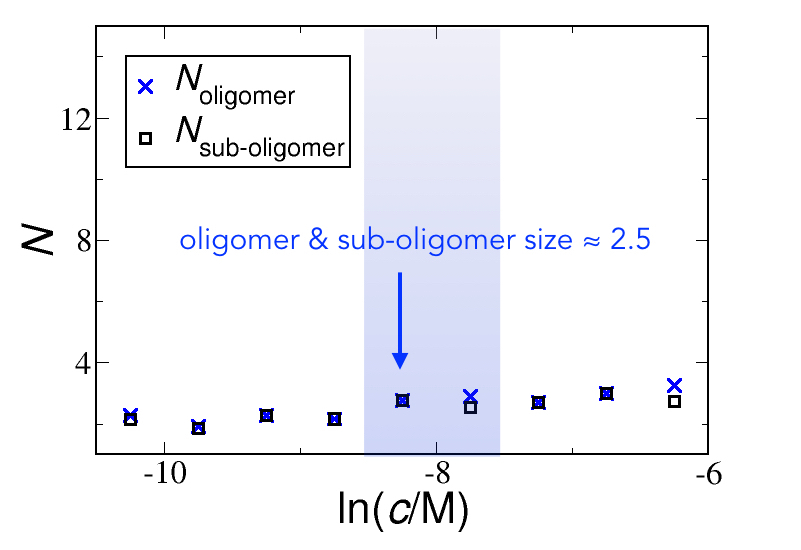}}
\caption{Kinetics of nucleation via oligomers from Monte Carlo simulations. (a) The rate of nucleation versus protein concentration measured in simulations, for proteins with low propensity for $\beta$-sheet ($\Delta \mu_{s\beta}=20\textrm{k}T$, red circles), and high $\beta$-sheet propensity ($\Delta \mu_{s\beta}=10\textrm{k}T$, blue crosses). (b) The concentration of free monomers in solution versus the total protein concentration. The solid line has the slope of 1, and the dashed line indicates the critical micelle concentrations. (c) The average size of the overall nucleating oligomer (red circles) for the low $\beta$-propensity protein ($\Delta \mu_{s\beta}=20\textrm{k}T$ from (a)), and the average size of the sub-oligomer participating in the conversion step (black circles) . (d) The average sizes of the overall nucleating oligomer (blue crosses) and the sub-oligomer participating in the conversion step (black squares) for the high $\beta$-propensity protein ($\Delta \mu_{s\beta}=10\textrm{k}T$ from (a)).}
\end{figure*}
\subsection{Control of kinetics of amyloid nucleation}
In what follows, we explore the factors that control the kinetics of amyloid nucleation, or in general,  nucleation via oligomers which includes a conformational change of the nucleating molecule. As previously discussed~\cite{frenkel2014}, the free energy barrier for nucleation with such two slow degrees of freedom is a trade-off between two opposing effects. At constant protein concentration, the probability of protein conversion from a soluble into the $\beta$-sheet prone state increases with the increase in the cluster size $N$. At the same time, the probability of formation of an oligomer of the size $N$, given by $c(N)$, decreases with $N$.  Hence, it follows that there is an intermediate cluster size that is optimal for nucleation at constant protein concentration, and this optimal cluster size increases with increasing protein concentration. In our simulations we can measure the conversion probability of a protein within an oligomer of the given size per MC step,\cite{footnote_PC} $P_c(N)$,  as shown in Fig. 4(a). We can also separately measure the steady-state concentration of oligomers at a certain protein concentration,  $c(N)$, from simulations under the situation when proteins are not allowed to convert into the $\beta$-prone state, as shown in Fig. 4(b). \\
Clearly, the conversion probability initially increases with increasing oligomer size (Fig. 4(a)). The reason for this observation is two-fold: firstly, larger oligomers contain a larger number of proteins available for conversion; secondly and more importantly, larger oligomers can have more possible binding partners to energetically stabilise the unfavourable `$s$'' $\rightarrow$ ``$\beta$'' conversion. The slope of this function $P_c(N)$ decreases abruptly at some point, after which $P_c(N)$ saturates to a plateau.
On the other hand, the oligomer concentration $c(N)$ decreases with increasing oligomer size (Fig. 4(b)). The probability of nucleation per MC step of a cluster of size $N$ should then be given by the product $P_c(N) \cdot c(N)$. Fig. 4(c) shows this product versus the oligomer size for five different protein concentrations. The nucleation probability per MC step at a certain protein concentration, given by $P_c(N) \cdot c(N)$, clearly exhibits a maximum, which corresponds to the most probable oligomer size for nucleation, $N^*$. It essential to notice that this oligomer size $N^*$ does not correspond to the most probable oligomer size observed in the system, which is $N=2$ (Fig. 4(b)).\\
The overall rate of nucleation should then depend, up to the prefactor, on the product of the two contributions, summed over all possible cluster sizes:
\begin{equation}
r \sim \sum_{N=1}^{\infty}  P_c(N) \cdot c(N).
\label{Eq_rate}
\end{equation}

\begin{figure*}[t!]
\centering
\subfigure[]{\includegraphics[width=0.45\textwidth]{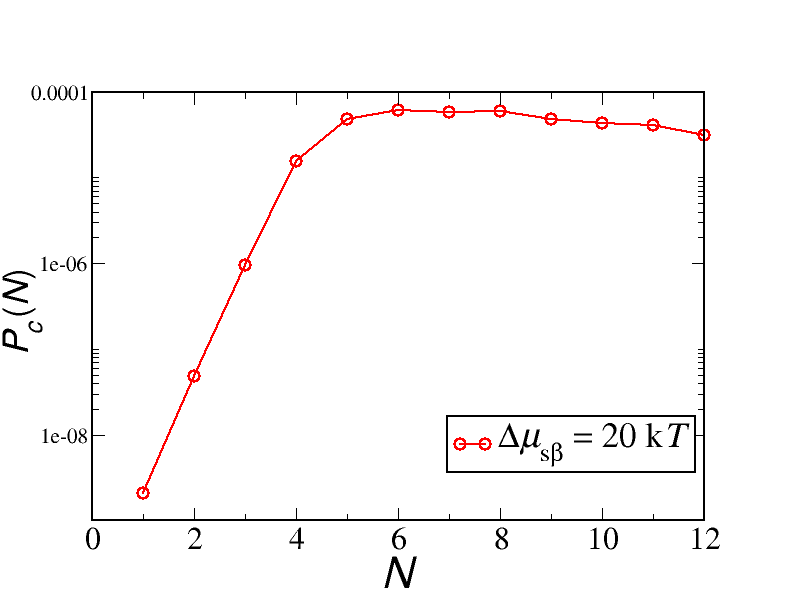}}
\subfigure[]{\includegraphics[width=0.45\textwidth]{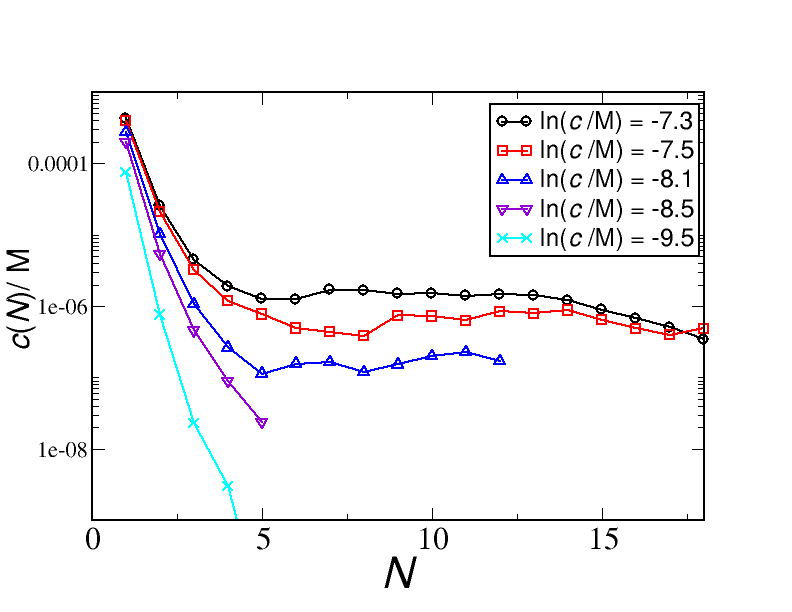}}
\subfigure[]{\includegraphics[width=0.45\textwidth]{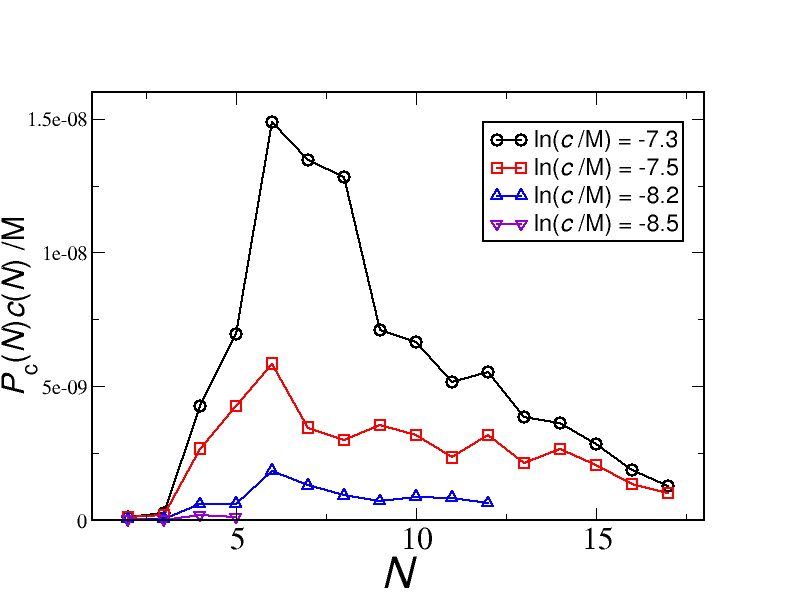}}
\subfigure[]{\includegraphics[width=0.45\textwidth]{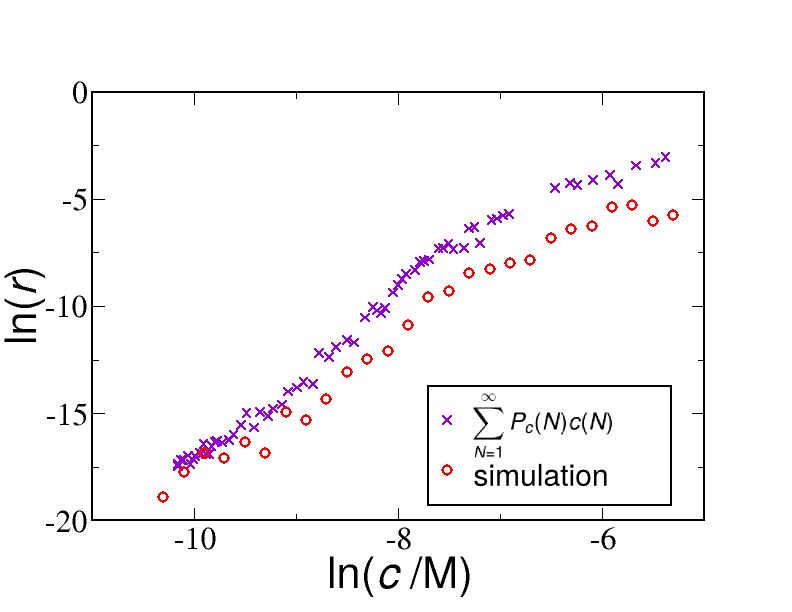}}
\caption{(a) The calculated probability of ``$s$'' $\rightarrow$ ``$\beta$'' conversion of a single protein within an oligomer of the size $N$, for the protein with a low $\beta$-propensity, $\Delta \mu_{s\beta}=20\textrm{k}T$, per one MC step. (b) Steady-state concentration of ``$s$''-state oligomers of size $N$ for five different protein concentrations. (c) Product of a conversion probability within an oligomer of a size $N$ per MC step, from (a), with the concentration of the oligomers of size $N$, from (b). (d) The calculated rate: Sum of the concentrations of all oligomers multiplied by the corresponding protein conversion rates within an oligomer, across the concentration range, as in Eq.(\ref{Eq_rate}). The nucleation rate measured in simulations for the same set of parameters is shown for comparison (red circles), taken from Fig.3(a).}
\end{figure*}\label{Fig4}

Note that the physical dimensions or $r$ are expressed in units of concentration per MC step. Using Eq.(\ref{Eq_rate}), we calculated the rate of nucleation, $r$, for the protein with a low $\beta$-propensity $\Delta \mu_{s\beta}=20\textrm{k}T$ (Fig. 4(d)), and compared it to the corresponding nucleation rate measured directly in simulations. The comparison shows an excellent agreement, up to a prefactor, between the calculated rate and the rate measured in simulations, giving the same scaling exponent. These results indicate that the reaction order is not only governed by the size of the nucleating cluster, as it is in the case of classical nucleation, but also by the probability of conformational conversion within such a cluster, which is an additional slow degree of freedom in this nucleation process.
\\
The probability for the conformational conversion is in general governed by the free energy difference between the two conformational forms, given by $\Delta \mu_{s\beta}$ in our simulations, and the interactions between the species within the oligomer, given by $\epsilon_{ss}$ and $\epsilon_{s\beta}$ in our simulations. To test the hypotheses that the scaling exponent is controlled by the functional form of $P_c(N)$, we computed the conversion probability per MC step $P_c(N)$ for three proteins with different $\beta$-propensities, while keeping the interaction parameters unchanged, as shown in Fig. 5(a). Trivially, the conversion probability increases with increasing $\beta$-propensity. However, the onset of the plateau occurs at smaller oligomer sizes for proteins with higher $\beta$-propensities, as marked by arrows in Fig. 5(a). This saturation can be viewed as the average number of neighbouring proteins needed for the conversion to become sufficiently probable. This number is smaller for proteins with higher $\beta$-propensity, and will determine the value of the reaction order. Fig. 5(b) shows the calculated nucleation rate, following Eq.(\ref{Eq_rate}), for the three proteins with different $\beta$-propensities. Indeed, a decrease in the $\beta$-propensity leads to an increase in the reaction order. Hence the functional form of $P_c(N)$ has a profound effect onto the reaction order, where the latter can be interpreted as the number of interacting proteins within the cluster necessary to stabilize the conformational conversion.\\

\begin{figure*}[t!]
\centering
\subfigure[]{\includegraphics[width=0.45\textwidth]{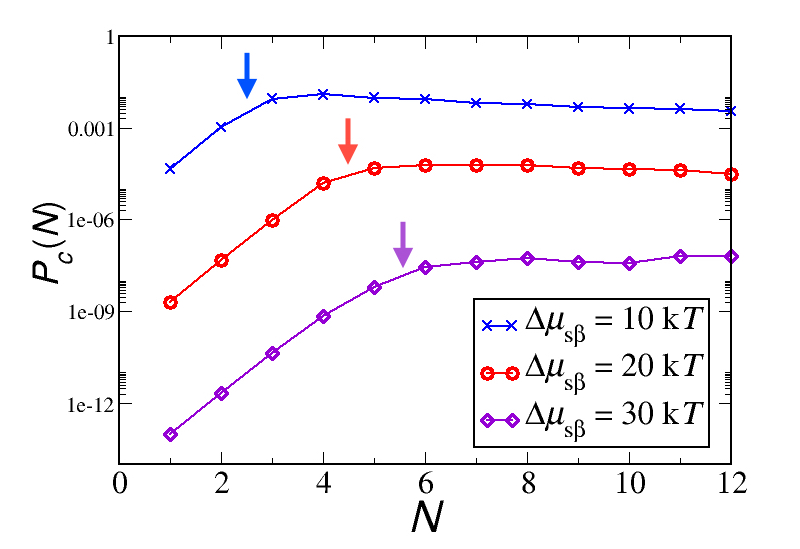}}
\subfigure[]{\includegraphics[width=0.45\textwidth]{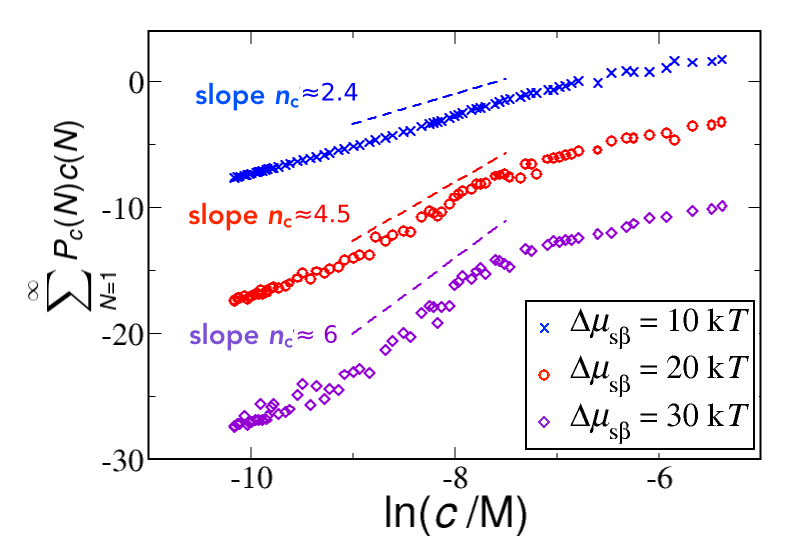}}
\caption{(a) The calculated probability of ``$s$'' $\rightarrow$ ``$\beta$'' conversion per MC step of a single protein within an oligomer of the size $N$, for three different $\beta$-propensities, from top to bottom: $\Delta \mu_{s\beta}=10\textrm{k}T$,  $\Delta \mu_{s\beta}=20\textrm{k}T$, and  $\Delta \mu_{s\beta}=30\textrm{k}T$. The arrows mark the respective saturation of the conversion probability. (b) The calculated rate, from Eq.(\ref{Eq_rate}), for three different $\beta$-propensities. }
\end{figure*}\label{Fig5}

\section*{Conclusions}
In this paper, we have studied the phenomenon of spontaneous nucleation in the context of protein filament formation, a process which has been discussed in the literature for over 50 years and has important implications in many areas of research. We have considered the phenomenon of spontaneous fibril formation first by assuming that the cluster size is the only relevant degree of freedom in the system. In this situation, the nucleation theorem states that the kinetic descriptors commonly measured in experiments, such as reaction orders and scaling exponents, relate in a direct way with the critical nucleus size$N^\ast$.\par 
We then introduced an additional slow coordinate in the nucleating system by allowing for a conformation conversion of the aggregating proteins. Using coarse-grained Monte-Carlo simulations, we probed the kinetics of this fibril nucleation process with two slow degrees of freedom. In the case of amyloid nucleation, this includes the aggregation number and the content of the fibrillar structure, the latter being characterised by the fraction of proteins in a $\beta$-sheet conformation. Our analysis showed that, in this case, the reaction order $n_c$ does not relate with the physical size of the aggregating oligomers, but that $n_c$ is modulated by the structural conversion within oligomers as the additional slow structural variable is describing the system. Under these circumstances, $n_c$ is found to correspond to the sub-cluster size that directly participates in the conversion and stabilizes the converted protein. These results provide direct practical insights into the interpretation of kinetic data of fibril formation.

\section*{Acknowledgments}
We acknowledge support from the Human Frontier Science Program and Emmanuel College (A.\v{S}), St John's and Peterhouse Colleges (T.C.T.M), the Swiss National Science Foundation (T.C.T.M.), the Biotechnology and Biological Sciences Research Council (T.P.J.K.), the Frances and Augustus Newman Foundation and  (T.P.J.K.), the European Research Council (T.C.T.M., T.P.J.K. and D.F), and the Engineering and Physical Sciences Research Council (D.F.).


\begin{thebibliography}{9}
\bibitem{footnote_PC}
The conversion probability per MC step of an oligomer of a size $N$ is measured by separately simulating the unconverted oligomer for $6 \cdot 10^{10}$ MC steps. At every MC step we computed the acceptance probability of conversion of every protein within the cluster, which was then averaged over the total number of MC steps. We expect that the oligomer will thoroughly sample its conformational space within the course of the simulation.
\end{thebibliography}
\end{document}